\documentclass[12pt,a4paper,twocolumn]{article}
\usepackage[left=2cm,right=2cm,top=2cm,bottom=2cm]{geometry}
\usepackage{authblk}
\usepackage{cite}

\usepackage[nottoc,other]{tocbibind}

\usepackage{sectsty}
\allsectionsfont{\centering}
\sectionfont{\fontsize{10}{15}\selectfont}
\subsectionfont{\fontsize{9}{12}\selectfont}
\subsubsectionfont{\fontsize{9}{12}\selectfont}
\usepackage[titletoc]{appendix}
\usepackage{listings,listingsutf8}
\usepackage{lmodern}
\usepackage{xcolor}
\usepackage{relsize} 
\usepackage{comment}
\usepackage{graphicx}
\graphicspath{ {images/} }
\usepackage{bm}
\usepackage[centerlast,scriptsize]{caption}
\usepackage{subcaption}
\usepackage{floatrow}
\usepackage{mathtools}
\usepackage{float}
\usepackage{amsfonts,amsmath,amssymb,amsthm}
\usepackage[hidelinks,colorlinks=false]{hyperref}
\usepackage[capitalise]{cleveref}

\usepackage{physics}
\usepackage{siunitx}
\usepackage{setspace}
\usepackage{ragged2e}
\usepackage{pdfpages}
\usepackage{multicol}

\usepackage{fancyhdr}

\title{\textbf{Chaos-based spread-spectrum communication system}}
\author[1]{\small \textbf{Christian Nwachioma}\thanks{christian.nwachioma@gmail.com}}
\affil[1]{\textbf{ CIDETEC, Instituto Polit\'ecnico Nacional, Mexico City 07700, Mexico} }
\providecommand{\keywords}[1]{\textbf{\textit{Index terms --- }} #1}
\date{}
\begin{document} 
	\onecolumn  
	\maketitle 	
	\thispagestyle{fancy}
	\normalsize
	\begin{abstract}
		\noindent Adaptive control is a control method that has an adaptation mechanism that reacts to model uncertainties. The control method is used to realized synchronization of a new chaotic system in a unidirectional master-slave topology. The master chaotic system and the slave system are adopted as transmitter and receiver, respectively for the purpose of secure communications. Both analog and digital designs are realized. The digital system is a cycle and bit accurate design having a system rate of $450MHz$. The design is targeted at an \textit{Artix-7 Nexys 4} FPGA. The transmitters are accordingly modulated by analog signals and fixed-point signals of different resolutions, sampling and frequencies. Although the adaptive controller tends to react on introduction of a modulating signal, we show that with a special detection mechanism or filtering including exponential smoothing, the choice of which depends on the nature of the modulating signal, it is possible to recover the modulating signal at the receiver. Moreover, chaos-based spectrum-spectrum communication systems are realized based on adaptive synchronization of the new chaotic system. Furthermore, in order to ascertain the robustness of the adaptive controller, the modulated signal is transmitted via an \textit{awgn} channel and the probability of error and bit-error-rate (BER) computed by sweeping across sets of SNR and noise power values in a Monte Carlo simulation. It turns out that the probabilities of error or error rates are reasonably low for effective communication through mediums with certain noise conditions. Hence, the adaptive controlled communication system can be robust.	
		\noindent\newline\newline\keywords{ Chaotic system; secure communications; adaptive control; analog and digital signals}
	\end{abstract}
\begin{figure}[h!]
	\includegraphics[width=0.8\linewidth]{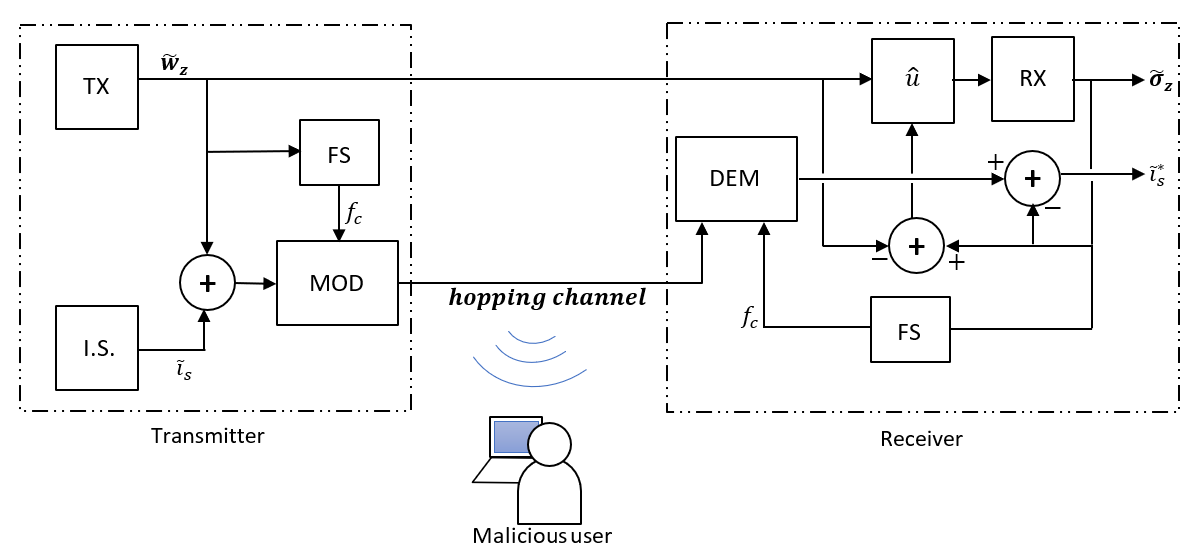}
	\caption{CFHSS communication system.}
	\label{secureCOMM_DSS}
\end{figure}

\begin{multicols}{2}
\begin{center}
	\section{INTRODUCTION}
	\label{ss}
\end{center}
In chaos-based spread-spectrum, the usual challenge is perfect synchronization of the transmitter and receiver. For a one-channel setup, it is not an easy task to achieve perfect synchronization of the chaotic transmitter-receiver system since the channel carries chaotic signal that has been modulated by some information signal. Besides, channel noise is unavoidably present in a real system. In\cite{15}, the workaround provided involves the setup of a dual-channel, where one channel is for the not-tampered chaotic signal and the other bears the information signal which has been encrypted using a sequence from the chaotic signal. At the receiver, the information is decrypted by a sequence from the corresponding synchronized signal. In the present report, we present two chaos-based spread-spectrum applications viz. chaos-based frequency-hopping spread-spectrum (CFHSS) and chaos-based direct-sequence spread-spectrum (CDSSS). CFHSS is based on a passband FM technology wherein the carrier frequency hops more or less chaotically by following a realtime sequence from the chaotic system. The CDSSS entails spreading an information signal of a slower rate than the system rate and a subsequent application of a correlator in order to have a communication. Finally, the same communication setups are considered in an \textit{awgn} channel and the error rates estimated. 
\end{multicols}
\begin{multicols}{2}
\begin{center}
	\section{CHAOS-BASED FHSS}
	\label{txss1}
\end{center}
In our chaos-based frequency-hopping spread-spectrum (CFHSS) application as \cref{secureCOMM_DSS} shows, there is a not-tampered chaotic signal for the controller through one of the channels and there is also the \textit{hopping channel}. In the \textit{hopping channel}, there is a continual frequency hopping through a modulator-demodulator passband network. The information signal $\tilde{i}_s$ is added to a chaotic signal and input into a passband frequency modulator (MOD). The carrier frequency $f_c$ of the modulator is derived from a frequency synthesizer (FS). The output of the synthesizer is dependent on quantized data of the chaotic signal. A modulo $m$ division is applied to the data, where $m-1$ is the number of frequency hopping channels. Since the system is sampled at $450MHz$, according Nyquist sampling theorem, we can choose the carrier frequency $f_c$ to be far less than half of the sample rate, i.e. $f_c < 1/2\times 450\enspace MHz$. For the purpose of numerical experimentation, we can assume that the low-power FM and FM bands\footnote{For real application purposes, carrier frequency bands may differ depending on country and available bands.} are free for the carrier frequency. In particular, let:
\small 
\begin{equation}
f_c\in \{58.6+1.4k, \enspace k = 1,..,m-1\}\enspace MHz,
\end{equation}
\normalsize
where $1.4\enspace MHz$ is the bandwidth of each channel. Assuming that $m=101$, \cref{channels} is a list of the channels, their frequency ranges and center frequencies. It should be noted that although the list of channels is sequential, their selection is not but based on the chaotic signals.
\end{multicols}
\begin{table}[h!]
	\caption{Frequency lookup table (LUT)}
	\label{channels}
	\begin{tabular}{ccc}
		\hline
		Channel indices, k & Frequency range (MHz) & Center frequency (MHz)\\
		\hline
		1	& 60.0 -- 61.4 &  60.7\\
		2 	& 61.4 -- 62.8 & 62.1\\
		.&.&.\\
		.&.&.\\
		99 & 197.2 -- 198.6 & 197.9\\
		100 & 198.6 -- 200.0 & 199.3\\
		\hline
	\end{tabular}
\end{table}
\begin{multicols}{2}
In order to recover the modulating signal, a corresponding passband frequency demodulator (DEM) is setup at the receiver. The carrier frequency $f_c$ of the demodulator has to be within the same channel bandwidth as the modulator. This can be made possible by feeding into the frequency synthesizer the corresponding synchronized chaotic signal. Although, the modulator and demodulator's frequencies are rapidly changing, they can always oscillate in near-enough synchrony at least within the channel's bandwidth provided synchronization of the chaotic transmitter-receiver system is realized. 

\begin{figure}[h!]
	\includegraphics[width=0.8\linewidth]{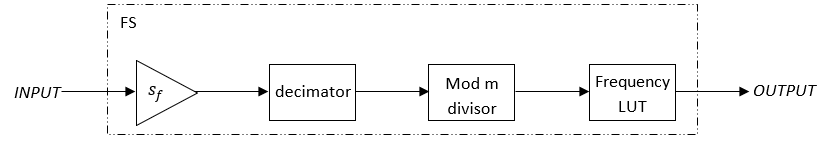}
	\caption{Frequency synthesizer (FS)}
	\label{Fsynthesizer}
\end{figure}

\begin{figure}[h!]
	\includegraphics[width=0.8\linewidth]{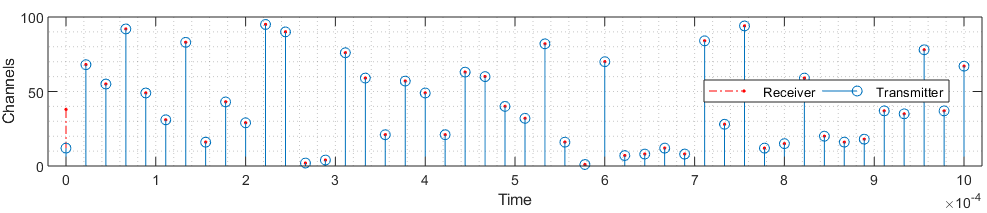}
	\caption{Chaos-based channel selection}
	\label{cselect}
\end{figure}

\begin{figure}[h!]
	\includegraphics[width=0.8\linewidth]{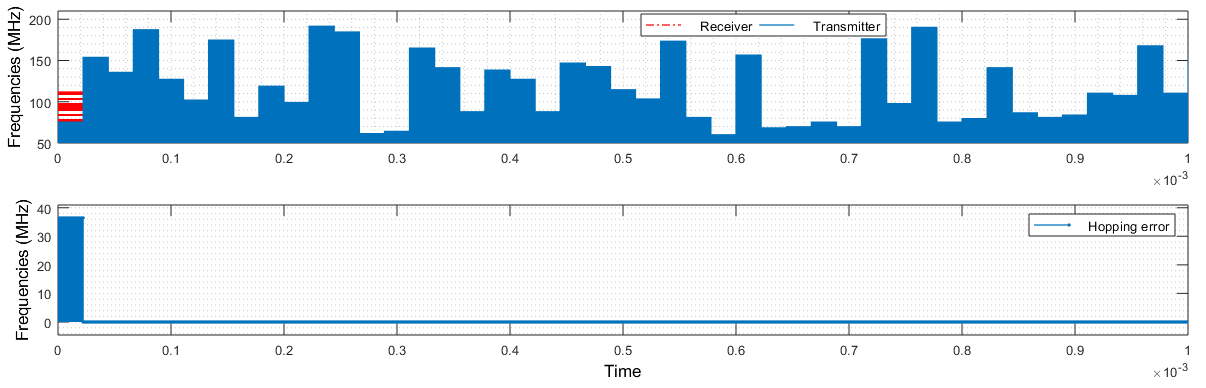}
	\caption{Chaos-based frequency hopping by the transmitter and receiver and the corresponding hopping error.}
	\label{freqHopping}
\end{figure}

\begin{figure}[h!]
	\includegraphics[width=0.8\linewidth]{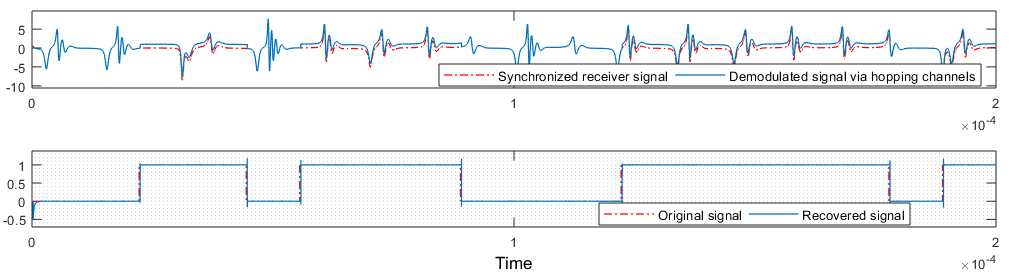}
	\caption{Recovery of a binary signal transmitted via a chaotically hopping channel.}
	\label{hoptxn}
\end{figure}

A frequency hopping as fast as the system sample rate will obviously prove problematic. Besides, this will lead to a channel lying close to the next. To solve this situation, a frequency hopping rate of $45kHz$, i.e. a factor of $10,000$ times slower than the system rate can be used. This can be realized by a frequency synthesizer as shown in \cref{Fsynthesizer}. The synthesizer receives a chaotic signal, upscales it by a factor $s_f$ to increase the resolution, a pulse generator of unit amplitude and with the appropriate period and duty cycle can be used to decimate the scaled chaotic signal. Then a modulo $m$ divisor and a floor rounding restrain the number of channels to $m-1$. Each channel serves as an index in the selection of a frequency from a lookup table. \cref{cselect} shows channel selection by the transmitter and receiver. Before synchronization, channels selected by the transmitter and receiver do not match in general. Subsequently, after synchronization is achieved, the selected channels coincide. Each channel corresponds to a frequency as presented in \cref{channels}. For any selected channel, the frequency is held for a period of $1/45000$ time unit, i.e. an inverse of the frequency hopping rate. Some of the transmitter and receiver hopping frequencies and corresponding error are shown \cref{freqHopping}. In order to compare the demodulated signal, which  is received via a hopping channel, with the corresponding synchronized chaotic signal, it should be noted that the latter lags the former by $100$ samples. Therefore, delaying the synchronized chaotic signal by $100$ samples and subtracting it from the demodulated signal, \cref{hoptxn}, shows that the original and recovered signals reasonably coincide. The little error, which may not be detrimental in digital application, is due to factors such as filtering, hopping, quantization and sampling (including resampling). 
In order to test the frequency hopping capability of our design, the \textit{SIMULINK} simulation time can be set to infinity, while the carrier frequency $f_c$ is updated on the fly using the \textit{MATLAB} function \textit{set\_param}.

The present scheme involving adaptive control has the advantage that keys are not shared for synchronization as in the conventional spread-spectrum application. Moreover, it has the advantage over the work in~\cite{25} since no random channel selection is necessary for exchange of chaotic seed values. The hopping of the frequencies provide security to the system as it limits the possibility of jamming, interception and malicious detection.
\end{multicols} 
\begin{figure}[h!]
	\includegraphics[width=0.8\linewidth]{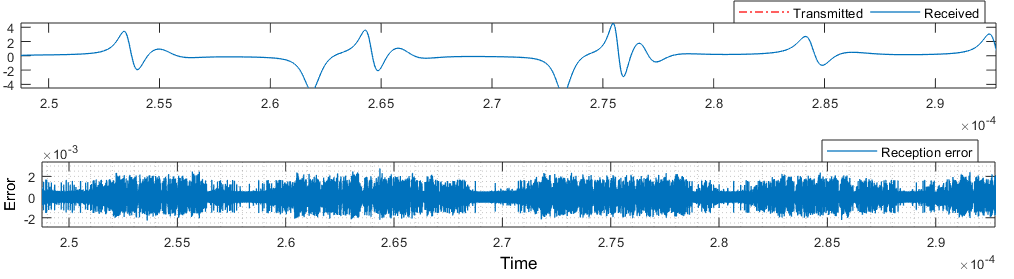}
	\caption{Modulated and demodulated signals by means of chaos-based frequency hopping and the corresponding reception error.}
	\label{hoptxn_error}
\end{figure}

\begin{figure}[h!]
	\includegraphics[width=0.8\linewidth]{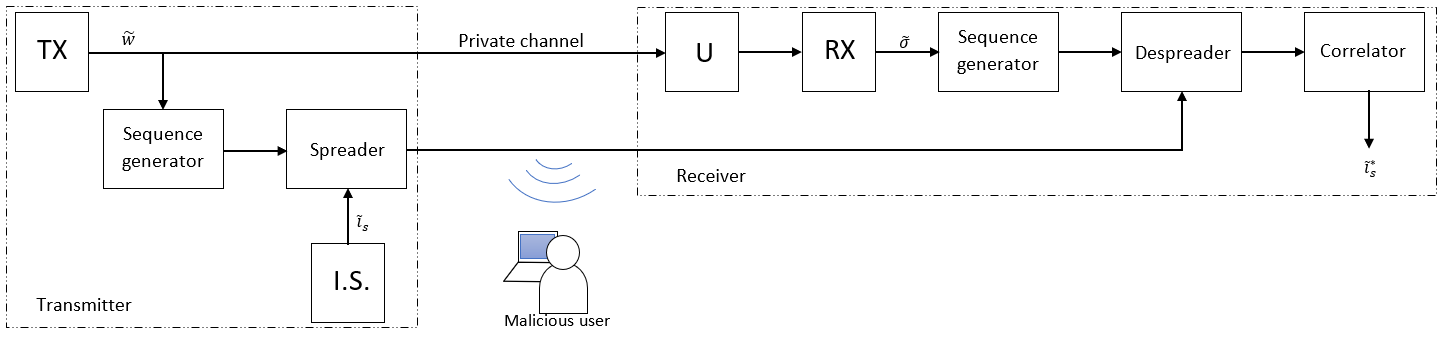}
	\caption{Dual channel chaos-based DSSS communication system.}
	\label{dsss}
\end{figure}

\begin{figure}[h!]
	\includegraphics[width=0.8\linewidth]{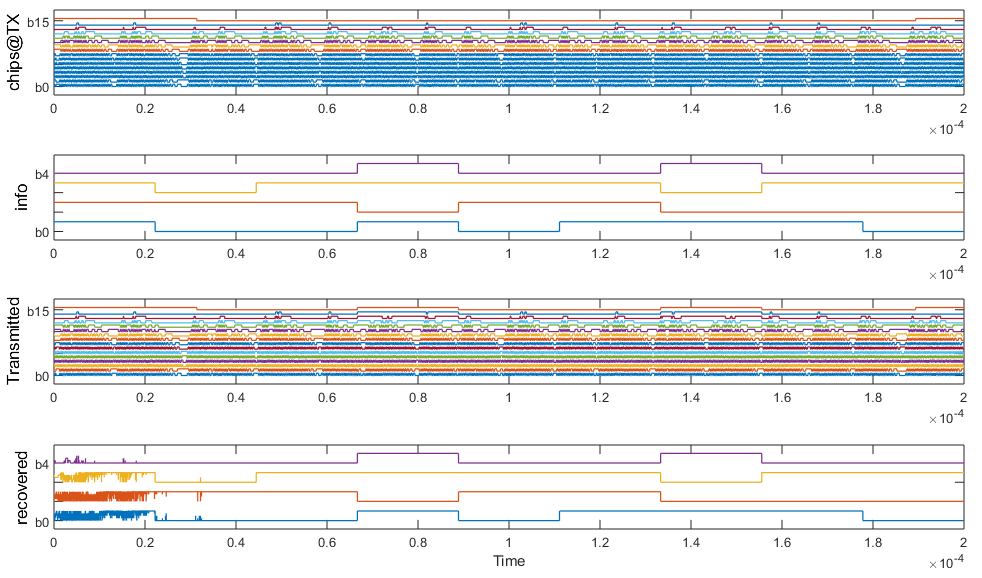}
	\caption{Chaos-based spreading and despreading of a 4-bit signal.}
	\label{dssseg}
\end{figure}

\begin{figure}[h!]
	\includegraphics[width=0.8\linewidth]{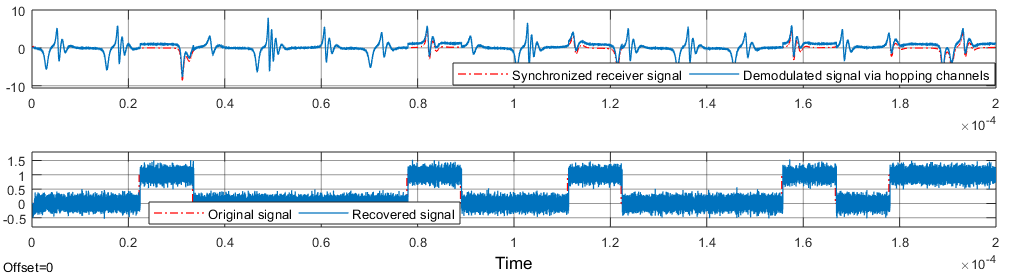}
	\caption{Chaos-based FHSS communication in the presence of lower power channel noise.}
	\label{hoptxn_noise}
\end{figure}

\begin{figure}[h!]
	\includegraphics[width = 0.8\linewidth]{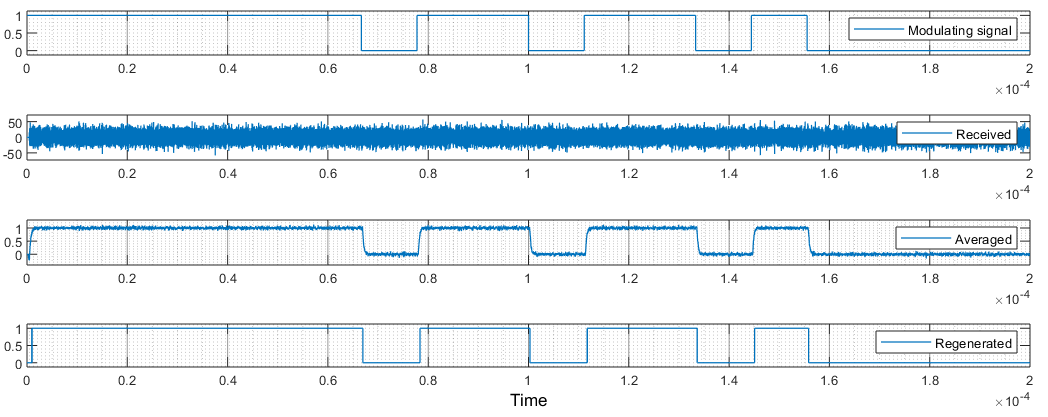}
	\caption{Transmission via a high-power awgn channel noise and recovery by exponential smoothing.}
	\label{windowfcn}
\end{figure}

\begin{figure}[h!]
	\includegraphics[width=0.8\linewidth]{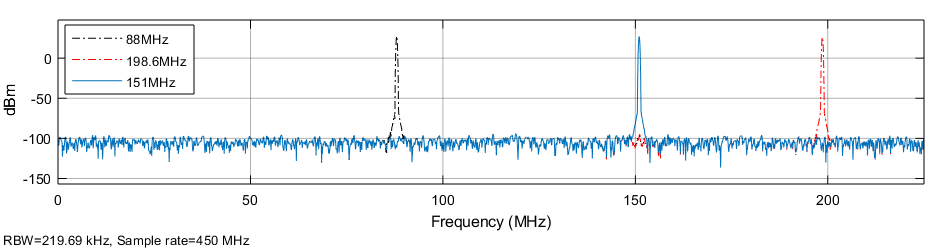}
	\caption{Spectrograph showing a few hopping frequencies amidst white Gaussian noise in the channel}
	\label{spectrogramHopping}
\end{figure}

\begin{figure}[h!]
	\includegraphics[width=0.8\linewidth]{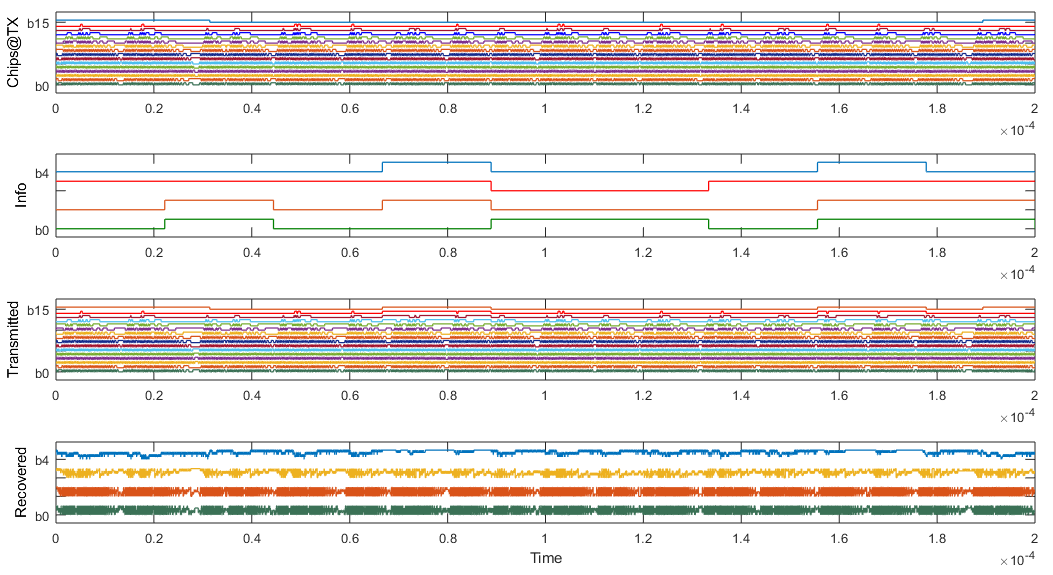}
	\caption{Chaos-based DSSS showing high sensitivity to channel noise.}
	\label{dssnoise}
\end{figure}

\begin{multicols}{2}
\begin{center}
	\section{CHAOS-BASED DSSS}
	\label{txss2}
\end{center}
The chip in a direct-sequence spread-spectrum (DSSS) application is often the \textit{m-sequence} or \textit{gold sequence}. In the present case, a sequence of bits is generated from the new chaotic transmitter for a CDSSS application (see \cref{dsss}). While signals in $\tilde{w}$ of the chaotic transmitter (\textit{TX}) are transmitted via a private channel for generating control signals, one of the signals can be input into a sequence generator, which creates an $m$-bit sequence at the system rate. The sequence can be used to spread the output of the information source (\textit{I.S.}), which has a much slower rate and $n$ bits $(n\le m)$. At the receiving end, the controller (\textit{U}) drives the receiver (\textit{RX}) to produce synchronized replicas of the transmitter's chaotic signals. The appropriate synchronized signal can be input into a sequence generator corresponding to the transmitter's. The generated sequence can be used to despread the received encoded signal. Since, the encoded data in general, do not have the same number of bits as the original information signal, a \textit{correlator} can be used to fully and correctly recover the information signal. For example, let us consider the spreading of a $4$-bit information signal, which has a rate of $45kHz$, i.e. $10,000$ times slower than the system rate. All $4$ bits of the information signal have the same rate. Since the chaotic transmitter is much faster than the information source and emits broadband signals, a well encoded signal that masks the information signal is produced for transmission. The information in the encoded data can be retrieved by operating on it with the appropriate despreading sequence. At the receiver, the corresponding synchronized chaotic signal is used to generate a despreading sequence. Applying the \textit{exclusive-or} formalism between the received encoded data and the despreading sequence results in a $16$-bit signal. In order to fully detect the information, a correlation is made that starts either from the \textit{MSB} or \textit{LSB} and summing successive $r = 16/4$ bits and then diving by $r$ according to eq.(4) in \textit{Part A}. This process should give only either a \textit{zero} or a \textit{one}. Wherever the despreading sequence does not match the spreading sequence, the result from \textit{despreading} and subsequent correlation will not agree with the original information data. This can be seen in \cref{dssseg} at the inception of stream of the recovered data. This happens because at those sample intervals, synchronization between the transmitter and receiver systems has not been reached. This is allowed in the current experiment for the purpose of illustration. In practice, we can allow a suitable latency before introducing the information signal. 
\end{multicols}
\begin{multicols}{2}
\begin{center}
	\section{PERFORMANCE ANALYSIS}
\end{center}
In the previous sections, the numerical experiments were conducted in an ideal environment. That is, the communication channels were devoid of noise or disturbances. Now, we shall introduce the white Gaussian noise of significant power level into the communication channel in order to test the robustness of the adaptive controller and detection mechanisms or filters.
\end{multicols}
\begin{multicols}{2}
\begin{center}
	\subsection{CFHSS IN A NOISY CHANNEL}
\end{center}
Since this approach employs dual channel, it is susceptible to higher level of noise. But a hopping channel can evade noise more than a fixed channel. Assuming a low power noise in the channel, it is possible to retain a good communication as \cref{hoptxn_noise} shows. The \textit{SNR} used in the current experiment is $50~dB$ in the $E_b/N_0$ mode and the noise level is $-30~dBm$. Of course, we can increase the signal amplitude with due consideration on hardware limitations, in order to retain communication in a channel with a higher noise power. Moreover, in the case of a more noisy channel, we can apply a window function such as exponential smoothing with a suitable forgetting factor in order to have a satisfactory visualization of the signal's signature. Then we can place a threshold to regenerate the binary signal without resorting to the commonly use filters. Thereby, recovering a non-degraded signal despite intense channel noise. For example, considering the same \textit{SNR} of $50~dB$ but with a much higher noise power level of $10~dBm$, the signature of the information signal is apparently completely lost in the noise (see \cref{windowfcn}). Applying an exponential smoothing, followed by a threshold, the information signal can be recovered as indicated in the figure. This shows that the CFHSS based on the adaptive controller together with some appropriate optimization techniques can prove useful in a real situation. Besides, the power spectral density\footnote{As the simulation runs and the frequency hops, only a single capture can be made at any instant. Robustness of the design can be better appreciated in a stream of moving pictures to be made available.} of signals present in the hopping channel is shown in \cref{spectrogramHopping}. We have used a \textit{Kaiser window} with a sidelobe attenuation up to $150~dB$ in order to have a distinct noise floor. 
\end{multicols}
\begin{multicols}{2}
\begin{center}
\subsection{CDSSS IN A NOISY CHANNEL}
\end{center}
The chaos-based DSSS communication system as presented is highly sensitive to channel noise. While the level of sensitivity can imply high security fidelity, it can prove difficult to realize using common communication channels especially for long distance communication as channel noise is inevitable. And the use of any form of averaging or filter affects the despreading sequence to the extent that the recovered signal remains scrambled. For a demonstration, using a high \textit{SNR} of $200~dB$ and a noise power as low as $-130~dBm$, the recovered signal yet appears nonsensical as can be seen in \cref{dssnoise}. In theory, we can increase further the \textit{SNR} or use a wider voltage difference to represent the logic \textit{highs} and \textit{lows} in order to recover a meaningful signal amidst channel noise. But the power specifications of the target device should be abided by. For example, today's FPGAs and some other embedded platforms have voltages no more than $3.3$ \textit{volts}. The present CDSSS channel noise problem can be investigated and reported elsewhere.
\end{multicols}
\begin{multicols}{2}
\begin{center}
	\section{CONCLUSION}
\end{center}
Adaptive control is a control method that has an adaptation mechanism that reacts to model uncertainties. While this feature might be desirable for stability of systems like aircraft and spacecraft, it unfortunately presents a serious challenge for communication purposes.
This is because the introduction of the information signal itself introduces deviations from the equilibrium and the adaptation mechanism by its nature reacts against the information signal in an attempt to return to the equilibrium. This has the tendency of distorting the control signal at the receiver thereby resulting in a recovered signal that might be different to the original information. However, despite this challenge, we show that it is possible to recover the information signal at the receiver. In some cases, this is made possible by additional techniques such as filtering, exponential smoothing and design of special detection mechanism.
This report discusses chaos-based spread-spectrum communication systems based on a new chaotic system. In particular, chaos-based frequency-hopping is used to provide secure communication using FM technology. The chaotically hopping carrier frequency of the passband FM modulator is shown to coincide with the demodulator's and a binary-valued bitstream is successful transmitted through the communication network. Moreover, a chaos-based DSSS is shown to be feasible using chips derived from the new chaotic transmitter and from the corresponding synchronized receiver. Furthermore, the designs are subjected to the real situation whereby channel noise is present. Despite transmission via an \textit{awgn} channel, most of the designs proved to be robust provided the \textit{SNR} is above a certain threshold or the noise power is below a certain threshold. Unfortunately, the chaos-based DSSS does not give a satisfactory result for an SNR as high as $200~db$, $-130~dBm$. This problem can be investigated further and reported elsewhere.
\end{multicols}
\begin{multicols}{2}
	\bibliographystyle{ieeetr}
	\bibliography{securecomm}
\end{multicols}
\end{document}